\documentclass[english]{paper}
\usepackage[T1]{fontenc}
\usepackage[latin9]{inputenc}
\usepackage{float}
\usepackage{graphicx}

\makeatletter

\providecommand{\tabularnewline}{\\}
\floatstyle{ruled}
\newfloat{algorithm}{tbp}{loa}
\providecommand{\algorithmname}{Algorithm}
\floatname{algorithm}{\protect\algorithmname}

\usepackage{algorithmic}

\makeatother

\usepackage{babel}
\begin{document}

\title{A nonenumerative algorithm to find the k longest (shortest) paths
in a DAG}

\author{Fatih Koçan}
\maketitle
\begin{abstract}
In this paper, we present a novel and efficient algorithm to find
the \emph{k} \emph{longest (shortest) paths} between sources and sinks
in a directed acyclic graph (DAG). The algorithm does not enumerate
paths therefore it is especially useful for very large \emph{k} values.
It is based on the Valued-Sum-of-Product (VSOP) tool, which is an
extension of Zero-suppressed Binary Decision Diagrams (ZBDDs). We
assessed the performance of this algorithm with a DAG model of a path-intensive
combinational circuit, \emph{viz.} c6288, that has $\sim10^{20}$
paths. We found that it took about 64 minutes to compute all paths
in this DAG along with their lengths.
\end{abstract}

\section{Introduction }

The classical problem of finding the shortest or longest paths in
a directed graph has been generalized to find the $k$ shortest or
$k$ longest paths. In the $k$ shortest path problem, the paths are
ranked in the increasing order of their lengths while in the $k$
longest path problem, the paths are ranked in the decreasing order
of their lengths. The $k$ shortest (longest) path problem is to find
$k$ shortest (longest) paths between two given vertices $s$ and
$t$ in a directed graph with non-negative edge weights for a positive
$k$ value. The shortest (longest) paths can be simple or with loops.
There are two well-known $k$ shortest nonsimple path finding algorithm.
One of them achieves $O(m+kn\: log\: n)$ \cite{Fox-JNM-75} and the
second one achieves $O(m+n\: log\: n+k)$ \cite{Eppstein:1999:FKS:299868.299886}
run times.

The problem of finding the $k$ shortest simple paths has been studied
in \cite{Hershberger:2007:FKS:1290672.1290682,Hoffman:1959:MSN:320998.321004,Law-MS-72,YenJY1,YenJY2,AzeSanSil-EJOR-93,BraSin-PEW-95}.
The best asymptotic performance has been obtained in \cite{YenJY1,YenJY2}
and it achieves $O(kn(m+n\: log\: n))$ worst-case run time. The work
in \cite{Hershberger:2007:FKS:1290672.1290682} outperforms the algorithms
with the best known results, esp. for large graphs. It can be noticed
that when $k$ is very large, such as $k=10^{15}$, the algorithms
become impractical.

There are also studies to compute the $k$ longest paths in DAGs \cite{21819,1586463,Ju:1991:ITI:127601.127729}..
All algorithms enumerative paths in the order of their lengths. In
this paper, we focus on finding $k$ longest paths in directed acyclic
graph (DAG) with any integer weights. Later, we modify the algorithm
to compute the $k$ shortest paths, the $k-th$ longest path, and
the $k-th$ shortest path. Our algorithm does not enumerate paths
and relies on the implicit calculation of paths and their lengths.
Also, we aim at very large $k$ values.

Our algorithm is inspired by the recent usage of the ZBDD tool in
path delay fault coverage (PDF) calculation \cite{PadmanabanMT03,KocanG05,KocanLS09}
and the availability of the Valued-Sum-of-Products (VSOP) tool \cite{MinatoVSOP06},
an extension to the ZBDD tool \cite{Minato93ZBDD}. The algorithms
are all \emph{nonenumerative} algorithms. The algorithms may encounter
memory overflow during their computations. In this case, the algorithms
would be run on partitioned graphs to avoid memory overflow that causes
the memory swapping, which in turn increase the run times \cite{KocanLS09}. 

The rest of the paper is organized as follows. Section II overviews
the VSOP tool. Section III introduces the $k$ longest (shortest)
path finding algorithm. Section IV gives the experimental result for
a path-intensive DAG. Finally, we conclude the paper.

\section{Valued-Sum-of-Product (VSOP) Tool based on ZBDDs }

A \emph{combinatorial item set} is a set of elements each of which
is a combination out of $n$ items. Zero-suppressed BDDs (ZBDDs) are
special type of BDDs that are designed for implicit representation
and efficient manipulation of combinatorial item sets \cite{Minato93ZBDD}.
The manipulation time is proportional to the size of the underlying
ZBDD. Therefore, ZBDDs utilize certain reduction rules to compactly
represent a set. The size of a ZBDD is also sensitive to the order
of items (i.e., variables) in a ZBDD. Many static and dynamic variable
ordering approaches are investigated for that purpose. We explain
a ZBDD with the help of an example.

\begin{figure}
\center\includegraphics[scale=0.8]{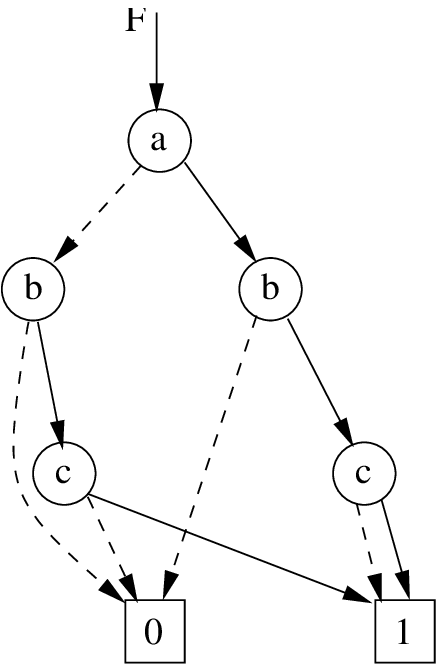}

\caption{ZBDD for $F=\left\{ abc,\: ab,\: bc\right\} $ \label{fig:example-zbdd}}
\end{figure}

Let $F=\left\{ abc,\: ab,\: bc\right\} $ be a combinatorial item
set. The set of items in $F$ are $\left\{ a,b,c\right\} $. Figure
\ref{fig:example-zbdd} illustrates the ZBDD for $F$. The items are
ordered as $a$ being the first, $b$ being the second, and $c$ being
the third. There are two types of edges: $0-$edges (dashed line)
and $1-$edge (solid line). Each path from the root to leaf $1$ corresponds
to a combination in the set. In a path, an solid (dashed) outgoing
edge from a node indicates that the respective item is included (excluded)
in (from) the combination. 

Valued-Sum-of-Product (VSOP) \cite{MinatoVSOP06,MinatoVSOP-report}
is an extension to ZBDD, and supports assignment of values to the
terms in a sum-of-products expression and enables efficient manipulation
of expressions. For example, $E=2ab+ac-3bd$ is a VSOP expression
where terms $ab$, $ac$, and $bd$ have $2$, $1$, and $-3$ values,
respectively. In VSOP algebra, addition and subtraction follow the
ordinary rule ($a+a=2a$) but the multiplication does not, and $a\times a=a$,
not $a^{2}.$ VSOP adopts the base ($-2$) binary encoding of numbers
and an integer number is decomposed into an $n-$digit vector of ZBDD
$\left\{ F_{0},F_{1},\ldots,F_{n-1}\right\} $. $F_{i}$ is a ZBDD
and stores the terms which have a $1$ in their $(i+1)^{th}$ digits
in their encodings. Specifically, $F_{0}$ is the set of odd-valued
terms, $F_{1}$ is the set of terms which have a $1$ in the second
position etc. 

The VSOP expression $F=4abc+5ab+3bc+a$ is represented in VSOP as
in Table \ref{tab:Encoding-of-values}. The encoding yields $F_{0}=\left\{ ab,bc,a\right\} $,
$F_{1}=\left\{ bc\right\} $, and $F_{2}=\left\{ abc,ab,bc\right\} $. 

\begin{table}
\caption{Encoding values of $F$'s terms\label{tab:Encoding-of-values}}

\center%
\begin{tabular}{|c|c|c|c|c|}
\hline 
$F$ & Value & $F_{2}$ & $F_{1}$ & $F_{0}$\tabularnewline
\hline 
\hline 
$abc$ & $4(100)$ & $1$ & $0$ & $0$\tabularnewline
\hline 
$ab$ & $5(101)$ & $1$ & $0$ & $1$\tabularnewline
\hline 
$bc$ & $3(111)$ & $1$ & $1$ & $1$\tabularnewline
\hline 
$a$ & $1(001)$ & $0$ & $0$ & $1$\tabularnewline
\hline 
\end{tabular}
\end{table}

The following VSOP operations are used in the following sections and
their details can be found in \cite{MinatoVSOP-report}.
\begin{itemize}
\item $E_{1}.Permit\left(E_{2}\right)$: Extract terms in $E_{1}$ each
of which is included in one of combinations in $E_{2}$.
\item $E_{1}.Restrict\left(E_{2}\right)$: Extract terms in $E_{1}$ each
of which includes one of combinations in $E_{2}$.
\item $E_{1}.TermsOP\left(E_{2}\right)$: Filter terms in $E_{1}$ \emph{``OP''}
to the constant term of $E_{2}$ where $OP=\left\{ EQ,\: NE,\: LE,\: LT,\: GT,\: GE\right\} $.
\item $E_{1}.OP\_Const(K)$: Filter terms in $E_{1}$ whose terms are ``OP''
to constant $K$ where $OP=\left\{ EQ,\: NE,\: LE,\: LT,\: GT,\: GE\right\} $.
\item $E_{1}\: op\: E_{2}$: Two expressions are subject to arithmetic or
logical operation $op$, where $OP=\left\{ +,-,*,/,\%,==,!=,<,<=,>,>=\right\} $. 
\item $E.CountTerms\left(\right)$: The number of minterms in expression
$E$.
\item $E.TotalVal\left(\right),\: E.MaxVal\left(\right),\: E.MinVal\left(\right)$:
Sum of the minterm values, maximum value in the set, and minimum value
in the set.
\item $I.GetInt()$: Convert VSOP value $I$ to integer.
\item $E.MinCover()$: Return a minimum-valued term in $E$.
\item $E.MaxCover()$: Return a maximum-valued term in $E$.
\end{itemize}
Let $F=4abc+5ab+3bc+a$ and $G=5ab-3bc$. Some operations on $F$
and $G$ and their outcomes are tabulated in Table \ref{tab:VSOP-examples-F}.
Comparison operations are better understood if we assume $0$ coefficients
for the nonexisting terms in $G$ i.e., $G=0abc+5ab-3bc+0a$.

\begin{table}
\caption{VSOP examples\label{tab:VSOP-examples-F}}

\center%
\begin{tabular}{|l|l|}
\hline 
Operation & Result\tabularnewline
\hline 
\hline 
$F.Restrict(a)$ & $4abc+5ab+a$\tabularnewline
\hline 
$F.Restrict(ab)$ & $4abc+5ab$\tabularnewline
\hline 
$F.Restrict(a+b)$ & $4abc+5ab+3bc+a$\tabularnewline
\hline 
$F.Permit(ab)$ & $5ab+a$\tabularnewline
\hline 
$F.Permit(abc)$ & $4abc+5ab+3bc+a$\tabularnewline
\hline 
$F.Permit(c)$ & $0$\tabularnewline
\hline 
$F.CountTerms()$ & $4$\tabularnewline
\hline 
$F.MaxVal()$ & $5$\tabularnewline
\hline 
$F.MinVal()$ & $1$\tabularnewline
\hline 
$F.TermsGE(3)$ & $4abc+5ab+3bc$\tabularnewline
\hline 
$F.TermsLT(3)$ & $a$\tabularnewline
\hline 
$F+G$ & $4abc+10ab+a$\tabularnewline
\hline 
$F-G$ & $4abc+6bc+a$\tabularnewline
\hline 
$F\times G$ & $5abc+30ab-9bc$\tabularnewline
\hline 
$F==G$ & $ab$\tabularnewline
\hline 
$F>G$ & $abc+bc+a$\tabularnewline
\hline 
$G>F$ & $0$\tabularnewline
\hline 
$F!=G$ & $abc+bc+a$\tabularnewline
\hline 
$F.MinCover()$ & $a$\tabularnewline
\hline 
$F.MaxCover()$ & $5ab$\tabularnewline
\hline 
\end{tabular}
\end{table}

\section{Algorithm }

This section introduces the algorithm that store all or selected paths
along with their lengths. Then, we select the k longest paths from
this path database. Later, we explain the k shortest path finding
algorithm. 

To find the k longest paths, we utilize a data-driven binary search
(DDBS) algorithm (Algo. \ref{alg:Data-driven-binary-search}). Since
sorting and index-based accessing to the paths in a VSOP is not possible,
the DDBS algorithm uses nonenumerative VSOP operators to find the
set of $k$ longest paths from the built path database. Also, in this
algorithm, when $\left\lfloor \frac{max+min}{min}\right\rfloor =min$,
the search to find the set is repeated for $min+1$ in the last step.
Otherwise, the algorithm would not terminate.

\begin{algorithm}[h]
\begin{algorithmic}
\REQUIRE $0 < K \le |\mathcal{L}|$
\STATE $ min \leftarrow \mathcal{L}.MinVal() $
\STATE $ max \leftarrow \mathcal{L}.MaxVal() $
\STATE $mid\_prev \leftarrow 0$
\WHILE{!done} 
\STATE $mid\_prev \leftarrow mid$
\STATE $ mid \leftarrow (min + max ) / 2 $
\IF{$CtoI\_EQ(mid,mid\_prev).GetInt()$}
\STATE $mid \leftarrow mid + 1$
\STATE $done \leftarrow 1$
\STATE $break$
\ENDIF
\STATE $ c1 \leftarrow (\mathcal{L}.EQ\_Const(mid)).CountTerms() $
\STATE $ c2 \leftarrow (\mathcal{L}.GT\_Const(mid)).CountTerms() $
\STATE $c3 \leftarrow c1+c2$
\STATE $flag \leftarrow c3.GT\_Const(K)) \: \&\& \: c2.LT\_Const(K)$
\IF{$(c3.EQ\_Const(K)) || flag)$}
\STATE $done \leftarrow 1$
\STATE $break$
\ELSIF{$(c3.LT\_Const(K)).GetInt()$}
\STATE  $max \leftarrow mid$
\STATE $continue$ 
\ELSIF{$(c3.GT\_Const(K)).GetInt()$}    
\STATE $min \leftarrow mid$
\STATE $continue$
\ENDIF
\ENDWHILE
\STATE return $\mathcal{L}.FilterThen(\mathcal{L}.GE\_Const(mid))$
\end{algorithmic}

\caption{$TopK(K,\mathcal{L})$\label{alg:Data-driven-binary-search}}
\end{algorithm}

We present an algorithm (Algo. \ref{alg:Paths-and-lengths}) to find
the $k$ longest paths with VSOP tool nonenumeratively. Algorithm
\ref{alg:Paths-and-lengths} starts from the source vertices and inductively
builds the path database in the topological order of vertices. At
each vertex, the algorithm builds the set0 of partial paths that end
at this node. The sum of sink nodes' paths is the set of all paths
along with their lengths. In the algorithm, we optionally utilize
early pruning of infeasible paths by calling $TopK()$ algorithm with
parameter $k$. 

In Algo. \ref{alg:Paths-and-lengths}, $\mathcal{L}$ and $\mathcal{L}_{i}$s
are VSOP expressions, and $\mathcal{L}_{i}==\mathcal{L}_{i}$ removes
the values of terms in a VSOP expression. For example, $\left(\left(2ab+3bc\right)==\left(2ab+3bc\right)\right)\Rightarrow ab+bc$.
The last loop takes the union of all paths that end at the sinks.
After that path queries can be performed on all paths using VSOP operations.
In our algorithm, we query the $k$ longest paths. Inductive building
of a set of paths is illustrated with the example below.

\begin{figure}
\center\includegraphics[scale=0.5]{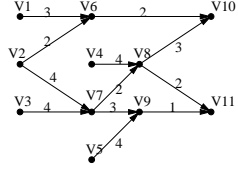}

\caption{A DAG with edge costs. \label{fig:Graph-model-of-c17}}
\end{figure}

\emph{Example:} In Fig. \ref{fig:Graph-model-of-c17}, $\mathcal{L}_{2}=0$,
and $\mathcal{L}_{3}=0$. $c_{i,j}$ is the length associated with
edge $e_{i,j}$. Thus, the partial paths at $v_{7}$ along with their
lengths are computed as follows. We initialize $\mathcal{L}_{7}=0$.
Since $v_{2}$ is a source vertex, $\mathcal{L}_{7}=\mathcal{L}_{7}+4\cdot v_{2}$.
Since $v_{3}$ is a source vertex, $\mathcal{L}_{7}=\mathcal{L}_{7}+4\cdot v_{3}$.
After that $\mathcal{L}_{7}=\mathcal{L}_{7}\cdot v_{7}=4\cdot v_{2}\cdot v_{7}+4\cdot v_{3}\cdot v_{7}$.
We compute compute $\mathcal{L}_{8}$ as follows. Since $v_{4}$ is
a source,$\mathcal{L}_{4}=0$. Thus, $\mathcal{L}_{8}=\mathcal{L}_{8}+4\cdot v_{4}$
and $\mathcal{L}_{8}=\mathcal{L}_{8}+\mathcal{L}_{7}+(v_{2}\cdot v_{7}+v_{3}\cdot v_{7})\cdot2=4\cdot v_{4}+6\cdot v_{2}\cdot v_{7}+6\cdot v_{3}\cdot v_{7}$.
Finally, $\mathcal{L}_{8}=\mathcal{L}_{8}\cdot v_{8}=4\cdot v_{4}\cdot v_{8}+6\cdot v_{2}\cdot v_{7}\cdot v_{8}+6\cdot v_{3}\cdot v_{7}\cdot v_{8}$.
We compute the partial paths at each vertex in topological order similarly
for all vertices.

\begin{algorithm}[H]
\begin{algorithmic}
\REQUIRE $V[1..N]$ : Topologically sorted vertices 
\FOR{$i \leftarrow 1; \:\: i \le N;\:\: i\leftarrow i+1 $}
\STATE $\mathcal{L}_i \leftarrow 0$
\FOR{each incidence vertex $v_j$ of $v_i$}
\IF{$v_j$ is a source}
\STATE $\mathcal{L}_i \leftarrow \mathcal{L}_i + c_{i,j} \cdot v_j$
\ELSE
\STATE $\mathcal{L}_i \leftarrow \mathcal{L}_i + \mathcal{L}_j + c_{i,j} \cdot (\mathcal{L}_j==\mathcal{L}_j)$
\ENDIF
\STATE $TopK(k,\mathcal{L}_i)$ -- early prune code 
\ENDFOR
\STATE $\mathcal{L}_i \leftarrow \mathcal{L}_i \cdot v_i$ 
\ENDFOR
\STATE $\mathcal{L} \leftarrow 0$
\FOR{each sink $v_i$}
\STATE $\mathcal{L} \leftarrow \mathcal{L} + \mathcal{L}_i$
\ENDFOR
\STATE $longest \leftarrow \mathcal{L}.MaxVal()$
\STATE return $TopK(k,\mathcal{L})$
\end{algorithmic}

\caption{Paths and their lengths: \textbf{ K-longest() }\label{alg:Paths-and-lengths}}
\end{algorithm}

By calling $TopK()$ algorithm after every partial path set calculation
at each vertex we can eliminate infeasible partial paths early from
the set of all partial paths at this vertex. A partial path that is
not in the top $K$ partial paths is an infeasible partial path. In
the final $k$ longest path set, there cannot be a path starting with
an infeasible partial path. Therefore, it is safe to remove them from
consideration. This early pruning of them would speed up the algorithm.

\subsection{Finding the \emph{k} shortest paths }

We modify the $TopK()$ algorithm (Algorithm \ref{alg:Data-driven-binary-search})
to find the $k$ shortest paths. All we need to:
\begin{enumerate}
\item change $GT$ into $LT$ in calculation of $c2$,
\item change $max\leftarrow mid$ to $min\leftarrow mid$, 
\item change $min\leftarrow mid$ to $max\leftarrow mid$, and
\item change $GE$ in the last statement to $LE$. 
\end{enumerate}
The modified algorithm is given in Algorithm \ref{alg:Data-driven-binary-search-1}.

\begin{algorithm}[h]
\begin{algorithmic}
\REQUIRE $0 < K \le |\mathcal{L}|$
\STATE $ min \leftarrow \mathcal{L}.MinVal() $
\STATE $ max \leftarrow \mathcal{L}.MaxVal() $
\STATE $mid\_prev \leftarrow 0$
\WHILE{!done} 
\STATE $mid\_prev \leftarrow mid$
\STATE $ mid \leftarrow (min + max ) / 2 $
\IF{$CtoI\_EQ(mid,mid\_prev).GetInt()$}
\STATE $mid \leftarrow mid + 1$
\STATE $done \leftarrow 1$
\STATE $break$
\ENDIF
\STATE $ c1 \leftarrow (\mathcal{L}.EQ\_Const(mid)).CountTerms() $
\STATE $ c2 \leftarrow (\mathcal{L}.LT\_Const(mid)).CountTerms() $
\STATE $c3 \leftarrow c1+c2$
\STATE $flag \leftarrow c3.GT\_Const(K)) \: \&\& \: c2.LT\_Const(K)$
\IF{$(c3.EQ\_Const(K)) || flag)$}
\STATE $done \leftarrow 1$
\STATE $break$
\ELSIF{$(c3.LT\_Const(K)).GetInt()$}
\STATE  $min \leftarrow mid$
\STATE $continue$ 
\ELSIF{$(c3.GT\_Const(K)).GetInt()$}    
\STATE $max \leftarrow mid$
\STATE $continue$
\ENDIF
\ENDWHILE
\STATE return $\mathcal{L}.FilterThen(\mathcal{L}.LE\_Const(mid))$
\end{algorithmic}

\caption{$TopK(K,\mathcal{L})$: shortest path version\label{alg:Data-driven-binary-search-1}}
\end{algorithm}

\subsection{Finding the \emph{k-th} longest and \emph{k-shortest} path}

After we find the $k$ longest paths, we can query the $k-th$ longest
path among them. Let $\mathcal{Z}$ be the set of $k$ longest paths.
$\mathcal{Z}.minCover()$ would return a minimum length path from
the set $\mathcal{Z},$ which is the $k-th$ longest path.

Similarly, we can find the $k$ shortest paths and query the $k-th$
shortest path among them. Let $\mathcal{Z}$ be the set of $k$ shortest
paths. $\mathcal{Z}.maxCover()$ would return a maximum length path
from the set $\mathcal{Z},$ which is the $k-th$ shortest path. 

Note that $k-th$ shortest or longest path may not be unique. In that
case, $TopK()$ algorithm may return more than $K$ paths as solution
set.

\section{Experimental Result }

We implemented the proposed algorithms and assessed their performances
using a DAG model of a path-intensive c6288 circuit benchmark \cite{ITC99}.
This benchmark has 32 source and 32 sink vertices with a total of
2448 vertices and 4800 edges. There are 9.89434$\times10^{19}$ paths
from sources to sinks in this benchmark. Experiment was performed
on an Intel-86-based 64-bit processor with two dual 2.66-GHz CPUs,
and 20-Gbyte RAM with the Linux operating system.

Table \ref{tab:c6288} tabulates the results for c6288. The edge weights
are set to a number between 1 and 10 that is generated randomly. We
calculate the number of used ZBDD nodes, time in minutes:seconds format,
percentage of memory usage, and the number of paths in the query of
top $K$ paths for various values of $K$. For example, Algorithm
\ref{alg:Paths-and-lengths} computes top $5x10^{5}$ paths in 6 minutes
and 34 seconds with 4.5\% of 20 GB memory. The algorithm allocates
20592319 ZBDD nodes and returns 572,976 longest paths. Note that we
do not find exactly $k$ paths when the $k-th$ path is not unique.

\begin{table}
\caption{c6288 \label{tab:c6288}}

\center%
\begin{tabular}{crrrr}
Top $K$ & Nodes & Time & Mem\% & Paths\tabularnewline
\hline 
$5x10^{5}$ & 20592319 & 6:34 & 4.5 & 572,976\tabularnewline
\hline 
$10^{6}$ & 26921861 & 8:33 & 4.6 & 1,485,955\tabularnewline
\hline 
$5x10^{6}$ & 36561672 & 12:19 & 8.9 & 6,561,113\tabularnewline
\hline 
$10^{7}$ & 44147879 & 15:31 & 9 & 16,629,545\tabularnewline
\hline 
All & 326301920 & 64:17 & 70.6 & $\sim10^{20}$\tabularnewline
\hline 
\end{tabular}
\end{table}

\section{Conclusion }

In this paper we introduced a novel algorithm based on VSOP and ZBDD
to extract the $k$ longest (shortest) paths and the $k-th$ longest
(shortest) path from a DAG. This algorithm outperforms existing algorithm
when the value of $k$ is very large since the run-times of the algorithms
become impractical for very large $k$ values. We assessed the performance
of our algorithm with a path-intensive DAG that has $\sim10^{20}$
paths. We were able to compute all paths and their lengths in about
64 minutes. 

The proposed algorithm can be used as a front end software to selected
critical paths for timing analysis. It can also used in the path delay
fault coverage calculation to enable delay-sensitive coverage calculation.

\bibliographystyle{plain}
\bibliography{longestpaths}

\end{document}